\documentstyle[12pt]{article}

\def\hybrid{\topmargin -20pt  \oddsidemargin 0pt
      \headheight 0pt   \headsep 0pt
      \textwidth 6.25in 
      \textheight 9.5in 
      \marginparwidth .875in
      \parskip 5pt plus 1pt   \jot = 1.5ex}

\hybrid

\begin{document}
\def\x{\times}
\def\beq{\begin{equation}}
\def\eeq{\end{equation}}
\def\beqa{\begin{eqnarray}}
\def\eeqa{\end{eqnarray}}
\def\L{ {\cal L}}
\def\C{ {\cal C}}
\def\N{ {\cal N}}
\def\calE{{\cal E}}
\def\lin{{\rm lin}}
\def\Tr{{\rm Tr}}
\def\cF{{\cal F}}
\def\cD{{\cal D}}
\def\modS{{S+\bar S}}
\def\mods{{s+\bar s}}
\newcommand{\Fg}[1]{{F}^{({#1})}}
\newcommand{\cFg}[1]{{\cal F}^{({#1})}}
\newcommand{\cFgc}[1]{{\cal F}^{({#1})\,{\rm cov}}}
\newcommand{\Fgc}[1]{{F}^{({#1})\,{\rm cov}}}
\def\mpl{m_{\rm Planck}}
\def\mxth{\mathsurround=0pt }
\def\xversim#1#2{\lower2.pt\vbox{\baselineskip0pt \lineskip-.5pt
x  \ialign{$\mxth#1\hfil##\hfil$\crcr#2\crcr\sim\crcr}}}
\def\simgr{\mathrel{\mathpalette\xversim >}}
\def\simle{\mathrel{\mathpalette\xversim <}}

\newcommand{\ms}[1]{\mbox{\scriptsize #1}}
\renewcommand{\a}{\alpha}
\renewcommand{\b}{\beta}
\renewcommand{\c}{\gamma}
\renewcommand{\d}{\delta}
\newcommand{\th}{\theta}
\newcommand{\TH}{\Theta}
\newcommand{\pa}{\partial}
\newcommand{\g}{\gamma}
\newcommand{\G}{\Gamma}
\newcommand{\A}{\Alpha}
\newcommand{\B}{\Beta}
\newcommand{\D}{\Delta}
\newcommand{\e}{\epsilon}
\newcommand{\E}{\Epsilon}
\newcommand{\z}{\zeta}
\newcommand{\Z}{\Zeta}
\newcommand{\k}{\kappa}
\newcommand{\K}{\Kappa}
\renewcommand{\l}{\lambda}
\renewcommand{\L}{\Lambda}
\newcommand{\m}{\mu}
\newcommand{\M}{\Mu}
\newcommand{\n}{\nu}
\newcommand{\X}{\Chi}
\newcommand{\R}{\Rho}
\newcommand{\s}{\sigma}
\renewcommand{\S}{\Sigma}
\renewcommand{\t}{\tau}
\newcommand{\T}{\Tau}
\newcommand{\y}{\upsilon}
\newcommand{\Y}{\upsilon}
\renewcommand{\o}{\omega}
\newcommand{\q}{\theta}
\newcommand{\h}{\eta}

\def\dota{ {\dot{\alpha}} }
\def\lag{Lagrangian}
\def\Kahler{K\"{a}hler}
\def\kahler{K\"{a}hler}
\def\A{ {\cal A}}
\def\C{ {\cal C}}
\def\F{{\cal F}}
\def\cL{ {\cal L}}

\def\R{ {\cal R}}
\def\x{ \times }
\def\beq{\begin{equation}}
\def\eeq{\end{equation}}
\def\beqa{\begin{eqnarray}}
\def\eeqa{\end{eqnarray}}

\sloppy
\newcommand{\be}{\begin{equation}}
\newcommand{\eq}{\end{equation}}
\newcommand{\ov}{\overline}
\newcommand{\un}{\underline}
\newcommand{\p}{\partial}
\newcommand{\la}{\langle}
\newcommand{\ra}{\rangle}
\newcommand{\bl}{\boldmath}
\newcommand{\ds}{\displaystyle}
\newcommand{\nl}{\newline}
\newcommand{\Nzahl}{{\bf N}  }
\newcommand{\zzahl}{ {\bf Z} }
\newcommand{\Zzahl}{ {\bf Z} }
\newcommand{\Qzahl}{ {\bf Q}  }
\newcommand{\Rzahl}{ {\bf R} }
\newcommand{\Czahl}{ {\bf C} }
\newcommand{\wt}{\widetilde}
\newcommand{\wh}{\widehat}
\newcommand{\fs}[1]{\mbox{\scriptsize \bf #1}}
\newcommand{\ft}[1]{\mbox{\tiny \bf #1}}
\newtheorem{satz}{Satz}[section]
\newenvironment{Satz}{\begin{satz} \sf}{\end{satz}}
\newtheorem{definition}{Definition}[section]
\newenvironment{Definition}{\begin{definition} \rm}{\end{definition}}
\newtheorem{bem}{Bemerkung}
\newenvironment{Bem}{\begin{bem} \rm}{\end{bem}}
\newtheorem{bsp}{Beispiel}
\newenvironment{Bsp}{\begin{bsp} \rm}{\end{bsp}}
\renewcommand{\arraystretch}{1.5}



\renewcommand{\thesection}{\arabic{section}}
\renewcommand{\theequation}{\thesection.\arabic{equation}}

\parindent0em

\def\S4{\frac{SO(4,2)}{SO(4) \otimes SO(2)}}
\def\P3{\frac{SO(3,2)}{SO(3) \otimes SO(2)}}
\def\MGd{\frac{SO(r,p)}{SO(r) \otimes SO(p)}}
\def\SOd{\frac{SO(r,2)}{SO(r) \otimes SO(2)}}
\def\SO2{\frac{SO(2,2)}{SO(2) \otimes SO(2)}}
\def\SUm{\frac{SU(n,m)}{SU(n) \otimes SU(m) \otimes U(1)}}
\def\SUS{\frac{SU(n,1)}{SU(n) \otimes U(1)}}
\def\SK{\frac{SU(2,1)}{SU(2) \otimes U(1)}}
\def\SU{\frac{ SU(1,1)}{U(1)}}

\begin{titlepage}
\begin{center}
\hfill CERN-TH/96-276\\
\hfill HUB-EP-96/53\\
\hfill SU-ITP-96-41\\
\hfill THU-96/34\\
\hfill {\tt hep-th/9610105}\\

\vskip .1in

{\bf CLASSICAL AND QUANTUM $N=2$ SUPERSYMMETRIC \\ BLACK HOLES}

\vskip .2in

{\bf Klaus Behrndt$^a$, Gabriel Lopes Cardoso$^b$, Bernard de Wit$^c$,\\ 
Renata Kallosh$^d$,
Dieter L\"ust$^a$, Thomas Mohaupt$^a$}\footnote{\mbox{email: \tt 
behrndt@qft2.physik.hu-berlin.de,
cardoso@surya11.cern.ch,
bdewit@fys.ruu.nl,} \\ \tt
kallosh@physics.stanford.edu,
                luest@qft1.physik.hu-berlin.de,
        mohaupt@qft2.physik.hu-berlin.de }
\\
\vskip 1cm

$^a${\em Humboldt-Universit\"at zu Berlin,
Institut f\"ur Physik, 
D-10115 Berlin, Germany}\\
$^b${\em Theory Division, CERN, CH-1211 Geneva 23, Switzerland}\\
$^c${\em Institute for Theoretical Physics, 
Utrecht University, 3508 TA Utrecht, Netherlands}\\
$^d${\em Physics Department, Stanford 
University, Stanford, CA 94305-4060, USA}

\vskip .1in

\end{center}

\vskip .2in

\begin{center} {\bf ABSTRACT } \end{center}
\begin{quotation}\noindent
We use heterotic/type-II prepotentials to study
quantum/classical black holes with half the $N=2, D=4$ 
supersymmetries unbroken.
We show that, in the case of heterotic string compactifications, the
perturbatively corrected entropy formula is given by the tree-level entropy
formula with the tree-level coupling constant replaced by the 
perturbative coupling constant.  In the case of type-II compactifications,
we display a new entropy/area formula associated with axion-free 
black-hole solutions, which depends on the electric and magnetic charges
as well as on certain 
topological data of Calabi--Yau three-folds, namely the intersection
numbers, the second Chern class and the Euler number
of the three-fold.  
We show that, for both heterotic and type-II theories, there is 
the possibility to relax the usual requirement of the 
non-vanishing of some of the charges and still have a finite entropy.

\end{quotation}
October 1996\\
\hfill CERN-TH/96-276\\
\end{titlepage}
\vfill
\eject

\newpage

\section{Introduction}

Recently there has been considerable progress in the 
understanding of microscopic and macroscopic properties of 
supersymmetric black holes in string theory. Using the 
Dirichlet-brane interpretation of type-II solitons, the 
microscopic entropy of certain stringy  black holes could be 
explicitly calculated \cite{Microscopic} in agreement with the 
macroscopic Bekenstein-Hawking entropy formula.       
In \cite{ExtrN2BHs} it was shown that, while the values of the 
moduli at spatial infinity are more or less arbitrary parameters, 
their values at the horizon are entirely fixed in terms of the 
(quantized) magnetic and electric charges of the black hole. 
In contradistinction with the black-hole mass, which is governed 
by the value of the  
central charge at infinity and thus depends on the charges and the 
moduli values at infinity, the entropy-area formula is given in 
terms of the central charge at the horizon \cite{FerKal1}. 
Because the  moduli are fixed at the horizon in terms of the 
charges, irrespective of their 
possible values at spatial infinity, the entropy-area is 
thus expressible in terms of the charges. This result is natural from the 
point of view that the entropy should follow from a counting of 
independent quantum-mechanical states, which seems to preclude 
any dependence on  
continuous parameters such as the moduli values at spatial 
infinity. In \cite{FerKal1} it was also shown that the central 
charge acquires a minimal value at the horizon and that the 
extremization of the central charge provides the specific 
moduli values at the horizon. 

The restricted behaviour at the horizon is related to the 
enhancement to full $N=2$ supersymmetry near the horizon, while 
globally the field configurations leave only half the 
supersymmetries unbroken (so that we are dealing with true BPS 
states). Thus the black holes can be regarded as solitonic 
solutions that interpolate between the maximally supersymmetric 
field configurations at spatial infinity and at the horizon. 
Particularly simple solutions, called double extreme black holes 
\cite{KSW},  
are given by those configurations where the moduli take constant 
values from the horizon up to spatial infinity. 

Using the extremization procedure of \cite{FerKal1}, the 
macroscopic entropy formulae 
for $N=4$ and $N=8$ extreme black hole solutions
\cite{D4N4Entr} were obtained in perfect agreement with the construction
of explicit black-hole solutions. The $N=4,8$ entropies are 
completely unique and they depend only on the quantized 
magnetic/electric charges and they are invariant under the 
perturbative and non-perturbative duality symmetries, such as 
$T$-duality, $S$-duality 
\cite{sduality} and string/string duality \cite{DuffKhuri,
HullTown,witt}.  

In four-dimensional $N=2$ string theories new features of black-hole
physics arise which destroy the uniqueness of the $N=2$ entropy formula.
In particular there exists a large number of different $N=2$ string vacua
so that the extreme black-hole solutions depend on the specific details
of the particular $N=2$ string model. Consequently the same 
features are present for the $N=2$ entropy formula. Nevertheless, 
the $N=2$ entropy, being proportional to the extremized $N=2$ 
central charge $Z$,  still depends on the quantized 
magnetic/electric charges, although 
the nature of the dependence is governed by the particular string
model. The $N=2$ central charge $Z$ and
the $N=2$ BPS spectrum can be directly calculated from
the $N=2$ holomorphic prepotential which describes the two-derivative 
couplings of the $N=2$ vector multiplets in the effective $N=2$ 
string action \cite{DWVP} (or, in a symplectic basis where the 
prepotential does not exist \cite{Ceresole}, from the symplectic 
sections). Therefore the  parameters of the prepotential of a
given $N=2$ string model determine the black-hole entropy as well as
the values of the scalar fields at the horizon.

Depending on whether one is discussing heterotic or type-II $N=2$
string vacua, the parameters of the prepotential have a rather different
interpretation.
To be more specific, let us first consider
four-dimensional $N=2$ heterotic string compactifications
on $K3\times T_2$, where the number of vector multiplets $N_V$
(not counting the graviphoton), the number of hypermultiplets $N_H$ and the
couplings are specified by a particular choice of the
$SU(2)$ instanton gauge bundle.
The  classical prepotential 
is completely universal and corresponds to a scalar non-linear
$\sigma$-model 
based on the coset space ${SU(1,1)\over U(1)}\otimes {SO(2,N_V-1)
\over SO(2)\times SO(N_V-1)}$.
Extremizing the corresponding central charge $Z$ the classical
$N=2$ black hole entropy and the
moduli on the horizon have been computed explicitly \cite{KSW,BKRSW,CLM},
and the result agrees with the truncated $N=4$ formulae.

Since in heterotic $N=2$ string compactifications the
dilaton field $S$ can be described by a
vector multiplet, the heterotic prepotential receives 
perturbative corrections only at the one-loop level \cite{DKLL,
AFGNT}; in addition there are non-perturbative contributions.
The heterotic one-loop corrections
to the prepotential, being independent of the dilaton $S$,
split into a cubic polynomial, a constant term and  
an infinite series of terms which are exponentially suppressed
in the decompactification limit of large moduli fields.
It is an interesting observation that the coefficients
of the exponential terms are given in terms of $q$-expansion
coefficients of certain modular forms as explicitly shown for
models with $N_V=3,4$ in \cite{HM,CCLM,CCL}. 
Thus, the one-loop black hole solutions are determined by an infinite
set of integer numbers; hence the extremization problem of the corresponding
one-loop central charges is very involved and
difficult. Nevertheless, we are able to derive a simple formula for
the black-hole entropy in terms of the heterotic string-coupling
and the target-space duality-invariant inner product of the
charges, which holds to all orders in perturbation theory. This
formula does not depend explicitly on the values of the moduli
fields. At the horizon the  values of the moduli can be 
determined explicitly in certain cases when
neglecting all exponential terms in the large moduli limit. Hence
new quantum features of black holes already
become important when considering only cubic corrections to
the classical prepotential.

It is well established 
that the $N=2$ heterotic string on $K3\times T_2$
is dual to type-IIA (IIB) compactification on a suitably chosen
Calabi--Yau three-fold \cite{KV,FHSV,AspLou}.
In fact, it was shown \cite{KV,KLT,KLM,AGNT,CCLM,CCL} 
for  models with $N_V=3,4$ that the type-IIA and
heterotic prepotential agree in heterotic weak-coupling limit.
On the type-IIA side the $N=2$ prepotential 
of the K\"ahler class moduli is completely classical
because the type-II dilaton corresponds to a hypermultiplet and has
no couplings to the vector fields. More specifically, 
the cubic couplings of the type-IIA prepotential are determined
by the topological intersection numbers of the corresponding Calabi--Yau
space; the coefficients of the exponential terms are given in
terms of the rational Calabi--Yau instanton numbers.
In this paper we focus on the limit of large K\"ahler-class moduli,
i.e. we will discuss the influence of the classical intersection
numbers $C_{ABC}$, as well as terms constant and linear in the moduli, 
on the Calabi--Yau black-hole solutions. The linear ones are 
related to the second Chern class of the three-fold and the 
constant one is related to the Euler characteristic \cite{HKTY1}. 
Hence we find new entropy formulae which depend only on the 
magnetic/electric charges and topological data on the 
Calabi--Yau manifold.

Our paper is organized as follows.
In the next section we will briefly introduce the $N=2$ vector couplings
and the $N=2$ central charge in terms of the $N=2$ prepotential.
We will recall the structure of the
prepotentials in four-dimensional $N=2$ heterotic and type-IIA string
vacua, and also their relations via heterotic/typeII string-string
duality.
In section~3 we show that there is a rather elegant and simple way
to find the solutions of the  extremization  problem of the $N=2$
central charge, which can be used 
to compute the values of the moduli on the black-hole horizon and the
black-hole entropy as a function of the quantized electric/magnetic
charges. While these solutions cannot be determined in full
generality, we can generally prove a simple formula for the entropy for
perturbative heterotic vacua, as a product of the inverse square of the
perturbative  string-coupling constant (which itself depends on
the electric/magnetic charges)  and the target-space
duality-invariant inner product of the charges. 
A particular class of solutions that can generally be evaluated
for cubic prepotentials, is the class of non-axionic
black-holes. 
This result covers the type-IIA Calabi--Yau black-hole entropy 
in case of small contributions of the rational instanton configurations,
i.e., in the limit of large K\"ahler-class moduli. 
We will also discuss the influence of linear terms in the prepotential
on the black hole entropy.
In the Calabi--Yau  case these linear terms are related to 
the second Chern class of the three-fold \cite{HKTY1}. 
In section four we discuss the relation of our solution to intersecting
branes in higher dimensions and suggest their M-theory 
interpretation. In the last section we summarize
our results.


\section{The $N=2$ prepotential in heterotic and type-IIA string 
vacua}

\setcounter{equation}{0}

\subsection{General formulae}

The vector couplings of $N=2$ supersymmetric Yang-Mills theory
are encoded in a holomorphic function $F(X)$, where the $X$ 
denote the complex scalar fields of the  
vector supermultiplets. With local supersymmetry this function 
depends on one extra field, in order to incorporate the 
graviphoton. The theory can then be encoded in terms of a 
holomorphic function $F(X)$ which is homogeneous of second 
degree and depends on complex fields $X^I$ with $I=0,1,\ldots 
N_V$. Here $N_V$ counts the number of physical vector multiplets. 

The resulting {\em special geometry} \cite{DWVP,special} can be defined 
more abstractly in  
terms of a symplectic section $V$, also referred to as period 
vector: a $(2N_V+2)$-dimensional 
complex symplectic vector, expressed in terms of the holomorphic 
prepotential $F$ according to 
\beqa
V= \pmatrix{ X^I \cr F_J\cr} \;\;, \label{sympsection}
\eeqa
where $F_I=\partial F/\partial X^I$. The $N_V$ physical scalar fields
of this system parametrize 
an $N_V$-dimensional
complex hypersurface, defined by the condition that the section satisfies a 
constraint 
\beqa
\langle \bar V,V\rangle \equiv \bar V^{\rm T}\Omega V 
= -i ,\label{symconstr}
\eeqa 
with $\Omega$ the antisymmetric matrix
\beqa
\Omega = \pmatrix{ 0& {\bf 1} \cr -{\bf 1} &0 \cr }\,.
\eeqa
The embedding of this hypersurface can be described in terms of
$N_V$ complex coordinates $z^A$ 
($A=1,\dots ,N_V$) by letting the $X^I$ be proportional to
some holomorphic sections $X^I(z)$ of the complex projective space.
In terms of these 
sections the $X^I$ read
\beq
X^I = e^{{1\over 2}K(z,\bar z)}\,
X^I(z)\,,\label{section} 
\eeq
where $K(z,\bar z)$ is the K\"ahler potential, to be introduced 
below. 
In order to distinguish the sections $X^I(z)$ from the original quantities
$X^I$, we will always explicitly indicate their $z$-dependence.
The overall factor $\exp[{1\over2}K]$ is chosen such that the constraint
(\ref{symconstr}) is satisfied. Furthermore, by virtue of the 
homogeneity property of $F(X)$, we can extract an overall factor 
$\exp[{1\over 2}K]$ from the symplectic sections (\ref{sympsection}), so 
that we are left with a holomorphic symplectic section. Clearly 
this 
holomorphic section is only defined projectively, i.e., modulo 
multiplication by an arbitrary holomorphic function. On the 
K\"ahler potential these projective transformations act as 
K\"ahler transformations, while on the sections $V$ they act as 
phase transformations. 
 
The resulting geometry for the space of physical scalar fields
belonging to vector multiplets of an $N=2$ 
supergravity theory is a special K\"ahler geometry,
with a K\"ahler metric $g_{A\bar B}=\partial_A\partial_{\bar B}K(z,\bar z)$
following from a K\"ahler potential of the special 
form
\begin{equation}
K(z,\bar z)=
-\log\Big(i \bar X^I(\bar z)F_I ( X^I(z))-i X^I(z)
\bar F_I(\bar X^I(\bar z))\Big) .
\label{KP} 
\end{equation}
A convenient choice of inhomogeneous coordinates $z^A$
are the {\em special} coordinates, defined by 
\begin{equation}
X^0(z)=1\,,\quad X^A(z)= z^A\,,\qquad A=1,\ldots ,N_V.
\end{equation}
In this parameterization the K\"ahler potential can be written as 
\cite{SU} 
\begin{equation}
K(z,\bar z) = -\log\Big(2({\cal F}+ \bar{\cal F})-
           (z^A-\bar z^A)({\cal F}_A-\bar{\cal F}_A)\Big)\,,
\label{Kspecial}
\end{equation}
where ${\cal F}(z)=i(X^0)^{-2}F(X)$. 

We should point out that it is possible to rotate the basis 
specified by (\ref{sympsection}) by an $Sp(2N_V+2,{\bf Z})$ 
transformation in such a way that it is no longer possible to 
associate them to a holomorphic function \cite{Ceresole}. As long 
as all fundamental fields are electrically neutral (which is the 
case in the context of this paper), this is merely a technical 
problem, as one can always rotate back to the basis where a 
prepotential exists \cite{Leuven}. As shown in \cite{Ceresole} 
the supergravity  
Lagrangian can be expressed entirely in terms of the symplectic 
section $V$, without restricting its parameterization so as to  
correspond to a prepotential $F(X)$. 

The Lagrangian terms containing the kinetic energies of
the gauge fields are
\begin{equation}
4 \pi \, {\cal L}^{\rm gauge}\ =\ -{\textstyle{i\over 8}}
\left( {\cal N}_{IJ}\,F_{\mu\nu}^{+I}F^{+\mu\nu J}\
-\ \bar{\cal N}_{IJ}\,F_{\mu\nu}^{-I} F^{-\mu\nu J} \right)\,,
\end{equation}
where $F^{\pm I}_{\m\n}$ denote the selfdual and anti-selfdual 
field-strength components and 
${\cal N}_{IJ}(z,\bar z)$ is the field-dependent tensor that 
comprises the inverse gauge couplings
$g^{-2}_{IJ}= {i\over 16\pi}({\cal N}_{IJ}-\bar {\cal N}_{IJ})$
and the generalized $\theta$ angles
$\theta_{IJ}= \frac{\pi}{2} ({\cal N}_{IJ}+\bar {\cal N}_{IJ})$.

Now we define the tensors
$G^{\pm}_{\mu\nu I}$ as
\begin{equation}
G^+_{\mu\nu I}={\cal N}_{IJ}F^{+J}_{\mu\nu}\,,\quad G^-_{\mu\nu 
I}=\bar{\cal N}_{IJ}F^{-J}_{\mu\nu}\,, \label{defG}
\end{equation}
which describe the (generalized) electric displacement and magnetic fields. 
The set of Bianchi identities and equation of motion for the 
Abelian gauge fields
are invariant under the transformations
\beqa
F^{+I}_{\mu\nu}\longrightarrow \tilde F^{+I}_{\mu\nu}  &=& 
U^I{}_J\, F^{+J}_{\mu\nu}+Z^{IJ}\,G^+_{\mu\nu J}\,.\nonumber\\ 
G^+_{\mu\nu I}\longrightarrow \tilde G^+_{\mu\nu I} &=& V_I{}^J\, 
G^+_{\mu\nu J}+W_{IJ}\,F^{+J}_{\mu\nu}\,, \label{FGdual}
\eeqa      
where $U$, $V$, $W$ and $Z$ are constant, real,  $(N_V+1)\times(N_V+1)$
matrices, which have to satisfy the symplectic constraint
\begin{equation}
{\cal O}^{-1} = \Omega\, {\cal O}^{\rm T} \,\Omega^{-1} \qquad
\mbox{where} \quad {\cal O} =
\pmatrix{ U & Z \cr W  & V\cr }\;.
\label{spc}
\end{equation}
The target-space duality group $\Gamma$ is a certain subgroup
of $Sp(2N_V+2,{\bf Z})$. 
It follows that the magnetic/electric 
charge vector $Q= (p^I,q_J)$, defined by 
$(\oint F^I, \oint G_J ) = (2\pi p^I,2\pi q_J)$,  
transforms as a symplectic vector, where we stress that 
the identification of magnetic and electric charges is linked to  
the symplectic basis. Since 
$N=2$ supersymmetry relates the $X^I$ to the field strengths
$F^{+I}_{\m\n}$, while the $F_I$ are related to the $G^{+\m\n}_I$,
the period vector $V$ also transforms as a symplectic vector:
\begin{eqnarray}
\tilde X{}^I&=&U^I_{\ J}\,X^J + Z^{IJ}\,
F_J\, ,\nonumber\\
\tilde F{}_I&=&V_I{}^J\,F_J + W_{IJ}\,X^J \,.
\label{xxx}
\end{eqnarray}

Finally consider $N=2$ BPS states, whose masses are equal
to the central charge $Z$ of the $N=2$ supersymmetry algebra.
In terms of the magnetic/electric charges $Q$ and the period vector $V$
the BPS masses take the following form \cite{Ceresole}:
\beqa
M_{BPS}^2=|Z|^2=|\langle Q,V\rangle |^2=e^K|q_IX^I(z)-p^IF_I(z)|^2
=e^{K(z,\bar z)}\,|{\cal M}(z)|^2.\label{bpsmasses}
\eeqa
It follows that $M_{BPS}^2$ is invariant under symplectic transformations
(\ref{xxx}).

An example of a prepotential arising in string compactifications is given 
by the cubic prepotential 
\begin{equation}
F(X)=d_{ABC}{X^AX^BX^C\over X^0}\,,\label{dfunction}
\end{equation}
where $d_{ABC}$ are some real constants.
The corresponding K\"ahler potential is given by 
\begin{equation}
K(z,\bar z)=-\log \Big(- id_{ABC}\,(z-\bar 
z)^A(z-\bar z)^B(z-\bar z)^C \Big)  \, .
\end{equation}
In the case of heterotic string compactifications, both the classical
prepotential
as well as certain perturbative corrections to it are described by a
cubic prepotential of the type (\ref{dfunction}).
In the case of type IIA compactifications, the $d_{ABC}$ are just proportional
to the classical intersection numbers: $d_{ABC}=-{1\over 6}C_{ABC}$.


\subsection{The heterotic prepotential}

In the following, we will discuss a class
of heterotic 
$N=2$ models, obtained by compactifying
the $E_8 \times E_8$ string on $K3\times T_2$.  The  moduli $z^A$
($A=1,\dots, N_V$)
comprise the dilaton $S$, the two toroidal moduli $T$ and $U$
as well as Wilson lines $V^i$ ($i=1,\dots ,N_V-3$):
\begin{equation}
S= -i z^1\,,\quad T= -i z^2 \,,\quad 
U= -i z^3 \,,\quad V^i= -i z^{i+3}\;.
\label{modid}
\end{equation}
We will, in the following, 
collectively denote the moduli $T,U$ and $V^i$ by $T^a$, so that  
$a=2,\ldots,N_V$.
The generic unbroken Abelian gauge group $U(1)^{N_V+1}$  
depends on the  specific choice 
of $SU(2)$ bundles with instanton numbers $(d_1,d_2)=(12-n,12+n)$
when compactifying to six dimensions on $K3$
(see \cite{KV,CF,AFIQ} for details).
For example, for $n=0,1,2$, a complete Higgsing is possible which
leads to the three-parameter $S$-$T$-$U$ models with no 
Wilson-line moduli ($N_V=3$). 
It is, however, also possible to not completely Higgs away the six-dimensional
gauge group $E_7\times E_7$,
and for $n=0,1,2$ one obtains in this way
heterotic models with one Wilson-line modulus $V$. Here we have 
four vector multiplets, so that we are dealing with 
a four-parameter $S$-$T$-$U$-$V$ model ($N_V=4$).

For this class of models, the heterotic prepotential has the  form
\beqa
{\cal F}^{\rm het}= 
- S T^a\eta_{ab}T^b  + h(T^a) + 
f^{\rm NP}(e^{- 2\pi S},T^a) \;,
\label{hetprepot}
\eeqa
where
\beqa
T^a\eta_{ab}T^b=T^2T^3-\sum_{I=4}^{N_V}(T^I)^2, \qquad a,b=2,
\dots ,N_V \;. 
\label{defeta}
\eeqa
The dilaton $S$ is related to the tree-level coupling constant and to the
theta angle by $S=4 \pi/g^2 - i \theta/2 \pi$.
The first term in (\ref{hetprepot}) is the classical part of the 
heterotic prepotential, $h(T^a)$ denotes the one-loop contribution
and $f^{\rm NP}$ is the non-perturbative part, which is 
exponentially suppressed for small coupling. Note that the 
perturbative corrections are entirely due to one-loop effects, 
owing to nonrenormalization theorems. In the following we 
focus on the perturbative contributions.

The classical prepotential leads to the metric of the special
K\"ahler manifold ${SU(1,1)\over U(1)}\otimes {SO(2,N_V-1)\over 
SO(2)\times SO(N_V-1)}$ with corresponding tree-level K\"ahler potential
\beqa
K       &=& -\log \Bigl[(S + \bar S)\Bigr ]-
        \log\Bigl[ (T^a + \bar T^a)\eta_{ab}(T^b + \bar T^b)\Bigr].
\label{Ktree}
\eeqa
   
Due to the required embedding of the $T$-duality group
into the $N=2$ symplectic transformations, it follows \cite{DKLL,
AFGNT} that the heterotic one-loop prepotential $h(T^a)$ must 
obey well-defined transformation rules under this
group.
The function $h(T^a)$ leads to the following modified 
K\"ahler potential \cite{DKLL}, which represents the full 
perturbative contribution,
\beqa
K = -\log\lbrack (S +  \bar S )+V_{GS}(T^a , \bar T^a)\rbrack -
\log\Bigl[  (T^a + \bar T^a)\eta_{ab}(T^b + \bar T^b) \Bigr] ,
\label{Kloop}
\eeqa
where
\beqa
V_{GS} (T^a , \bar T^a)
= {2( h + \bar h) - (T^a + \bar T^a)(\partial_{T^a} h  +  
\partial_{\bar T^a}
\bar h)
\over  (T^a + \bar T^a)\eta_{ab}(T^b + \bar T^b)}
\label{GSfunction}
\eeqa
is the Green-Schwarz term \cite{HOLAN} describing the mixing
of the dilaton with the moduli $T^a$.
Note that the true perturbative coupling constant is given by
\beqa
{4 \pi\over g^2_{\rm pert}}= {\textstyle{1\over 2}}\Big(S + \bar 
S+V_{GS}(T^a , \bar T^a)\Big) 
\;.\label{oneloopc}
\eeqa

To be more specific, let us recall the precise form of the one-loop
prepotential $h(T^a)$
\cite{HM,CCLM,CCL}. For simplicity we limit the
discussion here to the models with $N_V=4$. 
Any of the $S$-$T$-$U$-$V$ models considered  can be simply
truncated to the three-parameter
$S$-$T$-$U$ model upon setting $V \rightarrow 0$.
For the class of $S$-$T$-$U$-$V$ models considered
here, the one-loop prepotential is given by
\beqa
h(T,U,V) = p_n(T,U,V) 
- c -  \frac{1}{4\pi^3}\sum_{k,l,b\in {\bf Z} \atop
(k,l,b)>0} c_n(4kl-b^2) Li_3({\bf e}[ikT+ilU+ibV]), 
\label{prepstuv}
\eeqa
where
$c=\frac{c_{n}(0) \zeta(3)}{8 \pi^{3}}$ and 
${\bf e}[x]={\rm exp} 2 \pi i x$.
The coefficients $c_n(4kl-b^2)$ are the expansion coefficients of 
particular Jacobi modular forms \cite{CCL}.
$p_n$ is a cubic polynomial of the form \cite{BKKM,LSTY,CCL}
\beqa
p_n(T,U,V)=-{\textstyle{1\over 3}}U^3-({\textstyle{4\over 
3}}+n)V^3+(1+{\textstyle{1\over 2}}n)UV^2+{\textstyle{1\over 2}}n
TV^2 \;\;.\label{cubicfa}
\eeqa
It is important to note that the expression (\ref{prepstuv}) is valid
in the specific Weyl chamber ${\rm  Re } \;T>{\rm Re}\;  U>2 {\rm Re} \; V$.

Now consider taking the limit $S,T,U,V\rightarrow \infty$ subject to
${\rm Re} \;  S> {\rm Re} \;  T> {\rm Re}\; U>2 {\rm Re}\; V$, 
in which all non-perturbative as well as perturbative
exponential terms are suppressed. Then, the heterotic prepotential
is simply given by the cubic polynomial 
${\cal F}^{\rm het} = 
-S(TU-V^2)+p_n(T,U,V)$. In the limit $V\rightarrow 0$, 
the perturbative
prepotential is completely universal. 
In the large-moduli limit $S,T,U\rightarrow
\infty$ (${\rm Re}\; S> {\rm Re}\; T >{\rm Re} \; U$),
which is the decompactification limit to
5 dimensions, the prepotential of
these three-parameter models
takes the form
\beqa 
{\cal F}^{\rm het}=-STU-{\textstyle{1\over 3}}U^3
-c,
\label{STUprep}
\eeqa
where $c=\ov{c}=\frac{c_{STU}(0) \zeta(3)}{8 \pi^{3}}$ and
$c_{STU}(kl) = \sum_{b} c_{n}(4kl - b^{2})$ (for any $n$).
Using (\ref{GSfunction}) 
it is straightforward to compute the one-loop term $V_{GS}$
which follows from the prepotential (\ref{STUprep}):
\beqa
V_{GS}(T,\bar{T}, U, \bar{U}) =   \frac{ 
(U+ \bar{U})^2}{
3(T + \bar{T})  } - \frac{4 c}{(T + \ov{T}) (U + \ov{U})}\;\;.
\eeqa


\subsection{The type-IIA prepotential}

As already mentioned, the prepotential in type-IIA Calabi--Yau 
compactifications, which depends on 
 the K\"ahler-class moduli $t^A$ ($A=1,\dots ,N_V=h_{1,1}$),
is of purely classical origin. Nevertheless it has the same structure
as the heterotic prepotential. In fact, for dual 
heterotic/type-IIA pairs 
the prepotentials are identical upon a suitable identification
of the K\"ahler-class moduli $t^A$ in terms of the heterotic fields
$S$ and $T^a$.

The type-IIA prepotential has the following general structure 
\cite{HKTY}: 
\beqa
{\cal F}^{\rm II} = -{1\over 6}C_{ABC}t^At^Bt^C 
- \frac{\chi \zeta(3)}{2 (2 \pi)^{3}}
+\frac{1}{(2 \pi)^3} \sum_{d_1,...,d_h}
n^r_{d_1,...,d_h} Li_3({\bf e} [i\sum_Ad_At^A]),
\label{ftype2}
\eeqa
where 
we  work inside the K\"ahler cone
$\sigma(K)=\{\sum_A t^A J_A | t^A > 0\}$. (The $J_A$ denote the
(1,1)-forms of the Calabi--Yau three-fold $M$, which generate the
cohomology group $H^{2}(M, {\bf R})$).
The cubic part of the type-IIA
prepotential is given in terms of the classical intersection
numbers $C_{ABC}$, 
whereas the coefficients $n^r_{d_1,...,d_h}$
of the exponential terms denote the rational instanton numbers of 
genus 0. Hence, in the limit of large K\"ahler class moduli, 
$t^A\rightarrow\infty$, 
only the classical part, related to the intersection numbers, survives.
Consider, for example, the four-parameter model based on the compactification
on the Calabi--Yau three-fold $P_{1,1,2,6,10}(20)$ with $h_{1,1}=4$ and 
Euler number $\chi=-372$ \cite{BKKM}.
The cubic intersection-number part of the type-IIA prepotential for this model
is given as
\beqa
&&-{\cal F}^{\rm II}_{\rm cubic}=t^2((t^1)^2+t^1t^3+4t^1t^4+2t^3t^4+3(t^4)^2)
+{\textstyle{4\over 3}}(t^1)^3+8(t^1)^2t^4 \nonumber\\
&&\hspace{6mm} +t^1(t^3)^2
+2(t^1)^2t^3 +8t^1t^3t^4
+2(t^3)^2t^4
+12t^1(t^4)^2 
+ 6t^3(t^4)^2+6(t^4)^3 .
\label{cubicf}
\eeqa
Some of the rational instanton numbers $n^r_{d_1,d_2,d_3,d_4}$
for this model are displayed in \cite{BKKM,CCL}.
This model is dual to the previously discussed
heterotic string compactification with $N_V=4$ and $n=2$ \cite{CCL}.
The necessary identification of heterotic and type-IIA moduli is given as
\beqa
t^1=U-2V,\qquad t^2= S-T,\qquad t^3=T-U,\qquad t^4=V  \;,
\label{coordin}
\eeqa
and one can explicitly check that for some instanton numbers and for
the Euler number
the relations
\beqa
n^r_{l+k,0,k,2l+2k+b} = - 2 c_2(4kl-b^2),  \;\;\;
\chi = 2 c_{2}(0)
\label{insmod}
\eeqa
are indeed satisfied. In addition, the cubic heterotic prepotential $p_2$
(cf. (\ref{cubicfa})) and the Calabi--Yau prepotential (cf. (\ref{cubicf}))
agree.

Finally, let us mention that, in a particular symplectic basis,
it is very convenient to add to the Calabi--Yau prepotential (\ref{ftype2})
a topological term which is determined by the second Chern class $c_2$ of
the three-fold $M$, which gives rise to terms linear in the 
K\"ahler-class moduli fields: 
\beqa
{\cal F}^{\rm II} = -{1\over 6}C_{ABC}t^At^Bt^C +\sum_A^h{c_2\cdot 
J_A\over24}\, t^A+ \cdots\;.
\label{ftype3}
\eeqa
The real numbers $c_{2} \cdot J_{A} = \int_{M} c_{2} \wedge J_{A}$ are
the expansion coefficents of $c_{2}$ with respect
to the basis $J_{A}^{*}$ of the cohomology group $H^{4}(M,{\bf R})$
which is dual to the basis $J_{A}$ of $H^{2}(M,{\bf R})$
(i.e. $\int_{M} J_{A}^{*} \wedge J_{B} = \delta_{AB}$).
It is clear from eq.(\ref{xxx}) that adding such a linear term to the
prepotential is equivalent to performing a symplectic transformation with
$U=V=1$, $Z=0$ and $W_{0A}={c_2\cdot J_A\over 24}$; hence it 
has just the effect of a constant shift in the theta angles
\cite{witteff}. In the next section we will see that adding such 
a topological linear 
term to the prepotential may have interesting effects
on the $N=2$ black hole entropy as a function of the 
magnetic/electric charges.

As an example we consider the three parameter model based on
the Calabi-Yau $P_{1,1,2,8,12}(24)$ with $h_{1,1}=3$ and 
$\chi=-480$, which is dual to the heterotic string 
compactification with $N_V=3$ and $n=2$. 
The corresponding prepotential can be simply obtained from 
(\ref{cubicf}) by setting $V=0$. 
Here the  linear topological term takes the form
\beqa
\sum_A^3{c_2\cdot J_A\over24}\,t^A={\textstyle{23\over 
6}}t^1+t^2+2t^3=S+T+{\textstyle{11\over 6}}U\;.
\label{linearterm}
\eeqa


\section{$N=2$ Supersymmetric black holes}
\setcounter{equation}{0}

In this section we consider extreme dyonic black holes in the 
context of $N=2$ supergravity. The fields corresponding to these 
black holes spatially 
interpolate between two maximally supersymmetric field 
configurations. One is 
the trivial flat space at spatial infinity, which allows constant 
values for the moduli fields. The other is the Bertotti-Robinson 
metric near the horizon, where the fields are restricted to 
(covariantly) constant moduli and graviphoton field strength (the 
latter is directly related to the value of the central charge at 
the horizon). The 
interpolating fields leave only half  
the supersymmetries invariant, so that we are dealing with true 
BPS states. For these black holes, the mass is equal to the 
central charge taken at spatial infinity, so that 
\be
M^2_{ADM}= \vert Z_\infty \vert^2 = e^{K(z,\bar z)}\,\vert {\cal 
M}(z) \vert^2 \Big\vert_{\infty} \;,
\eq
where the moduli fields $z$ are taken at spatial infinity. Hence the 
mass depends generically on the magnetic/electric charges and the 
asymptotic values of the moduli fields. 

Near the horizon the values of the moduli fields, and thus the 
value of the central charge, 
are strongly restricted by the presence of full $N=2$ supersymmetry.
In \cite{FerKal1} 
it was proved that this implies that the central charge becomes 
extremal on the horizon. The result of this is that one can 
express the values of the moduli at the horizon in terms of the  
magnetic/electric charges $p^I$ and $q_I$. The value of the 
central charge at the horizon is related to the Hawking-Bekenstein entropy,
\be
{{\cal S}\over \pi} =  \vert Z_{\rm hor}\vert^2\;, \label{entropy}
\eq
where we have conveniently adjusted the value of Newton's 
constant.  The area of the black hole, which equals four times 
the entropy, has an interpretation as the mass of the 
Bertotti-Robinson universe. The crucial observation here is that 
the entropy and related quantities depend only on the quantized 
magnetic/electric charges (with $N=2$ supersymmetry the nature of 
this dependence is governed by the particular string 
vacuum), while the mass of the black hole  
depends on the charges as well as on the asymptotic values of the 
moduli. The latter are, in principle, arbitrary parameters that 
do not depend on the charges and, when approaching the black 
hole, evolve according to a 
damped geodesic equation towards the fixed-point values at the 
horizon, which are given in terms of the charges.  

There exist so-called double extreme black holes, introduced in 
\cite{KSW}, for which the moduli remain constant away from the 
horizon. In that case the central charge remains constant and 
thus the black hole mass is equal to the Bertotti-Robinson mass. 
The moduli at spatial infinity take the same values as near the 
horizon, so that the black-hole mass itself is now also a 
function of the magnetic/electric charges. 
Consequently, for double extreme black holes we find that 
$M_{ADM}$ is a function of the $p^I$ and $q_I$.

In this section we study the extremization problem at the 
black-hole horizon to obtain the value of the moduli and the 
black-hole entropy as a function of the charges $p^I$ and $q_I$. 
We cast this problem in a convenient form, which can be 
formulated in
terms of a variational principle (cf. (\ref{potential})). This 
allows us to construct a variety of explicit solutions. Then, in
the second subsection, we consider the black-hole entropy for 
heterotic $N=2$ supersymmetric string compactifications to all orders
in string perturbation theory and derive a general formula for
the entropy. An important feature of this formula is its 
invariance under target-space duality. In the next subsection we 
consider so-called 
non-axionic black holes, where one can conveniently obtain 
explicit solutions. 
Finally in subsection 3.4 we consider the entropy for type-II
compactifications. 


\subsection{Extremization of the $N=2$ central charge}

Let us start and exhibit some features of the double extreme 
black holes, for which the moduli remain constant. The metric of 
these black holes is of the extreme Reissner-Nordstrom form with 
the mass equal to the   
square root of the area divided by $4\pi$. In isotropic 
coordinates the metric is 
\begin{equation} \label{4dmetric}
ds^2 =-  \left (1+ {\sqrt  {A / 4 \pi }\over r}\right)^{-2} dt^2 +  \left (1+
{\sqrt  {A / 4 \pi }\over r}\right)^{2}  d\vec x^2 \;\;.
\end{equation}
One can also present the metric as  ($\tilde r = r + \sqrt  {A / 4 \pi }$)
\begin{equation}
ds^2 =-  \left (1- {\sqrt  {A / 4 \pi }\over \tilde r}\right)^{2} 
dt^2 + \left (1- {\sqrt  {A / 4 \pi }\over \tilde r}\right)^{-2}  
d\tilde r^2 + \tilde r^2 d^2 \Omega \;\;.
\end{equation}
The mass is defined  via the large-$\tilde r$ expansion
\begin{equation}
g_{tt} = \Big(1 - {2M_{ADM}\over \tilde r} +\cdots \Big) 
\end{equation}
and the metric shows that
\begin{equation}
M_{ADM}= \sqrt  {A \over 4 \pi} \;\;.
\end{equation}
In this form  it is clear that the horizon is at $g_{tt}=0 \Longrightarrow
\tilde r =  \sqrt  {A / 4 \pi }$. Therefore the area of the horizon is indeed
given by
\begin{equation}
4\pi (\tilde r^2) _{\rm hor} = A \;\;.
\end{equation}

As discussed above, to obtain the value of the moduli at the 
horizon for extreme $N=2$ black holes, one can determine the extremal 
value of the central charge in moduli space. This implies that 
\begin{equation}
\partial_A |Z |= 0 \;.
\end{equation}       
These equations are difficult to solve in general.
They are, however, equivalent to the following set of equations
\cite{FerKal1} 
\beqa
\bar Z \, V - Z\,\bar V = i Q\, ,\label{simplcond}
\eeqa
where $Q$ is the magnetic/electric charge vector $Q=(p^I,q_J)$. 
The above relation is closely related to the fact that the field 
configurations are fully supersymmetric at the horizon. 
Here we note that these equations can be independently justified on
the basis of symplectic covariance. Assuming that the moduli near 
the horizon 
depend exclusively on the magnetic/electric charges and satisfy
equations of motions that transform in a well-defined way
under symplectic (duality) reparametrizations, the symplectic
period vector must be proportional to the symplectic charge
vector. As the period vector is complex and the charge vector is
real, there is a complex proportionality factor which must be a
symplectic invariant. Using (\ref{symconstr}) we derive that this
factor is precisely the central charge $Z$ and find the above
result (\ref{simplcond}). 

{From} (\ref{simplcond}) one can determine the period vector, which
is defined in terms of $N_V$ complex moduli. We do this by
reformulating the equation and the corresponding expression of
the black-hole entropy in terms of a variational principle. To do
this, we first introduce 
a new symplectic vector $\Pi$ by 
\be
\Pi =\pmatrix{Y^I\cr F_J(Y)\cr}\qquad  \mbox{where } \; 
Y^I\equiv  \bar Z\,X^I\,.\label{ydef}
\eq
Observe that $Y^I$  and thus the vector $\Pi$ is $U(1)$
invariant, so that it is not subject to K\"ahler
transformations. In terms of $\Pi$, (\ref{simplcond}) and 
(\ref{entropy}) turn into
\beqa
\Pi -\bar \Pi = iQ\,, \qquad  {{\cal S}\over \pi} = \vert Z_{\rm 
hor}\vert ^2 = i
\langle \bar \Pi, \Pi  \rangle\,.   \label{picond}
\eeqa
The equations (\ref{picond}) are governed by a variational principle 
associated with a `potential' 
\beqa
{\cal V} _Q(Y,\bar Y) \equiv  -i\langle \bar \Pi , \Pi 
\rangle  - \langle \bar \Pi +\Pi , Q\rangle\,. \label{potential}
\eeqa
${\cal V}_Q$ takes an extremal value whenever $Y$ and $\bar Y$ 
satisfy the first equation (\ref{picond}). This extremal value is given by 
the second expression 
(\ref{picond}) for the entropy. 
Using (\ref{KP}) the entropy can be also written as 
\beqa
 {{\cal S}\over \pi} = |Y^0|^2 \exp\Big[- K(z,\bar z)\Big] 
\bigg\vert_{\rm hor}\;.    \label{ADMKP}
\eeqa
where  $Y^0$ and the special coordinates  $z^A$ are evaluated at 
the horizon. 
 
Let us now consider the construction of solutions to the
equations (\ref{picond}). Written in components they read
\beqa
Y^I-\bar Y^I = i p^I \, ,\qquad F_I(Y) -\bar F_I(\bar Y) = i 
q_I\,.                \label{ycond}
\eeqa
To solve these equations it does not help to go to a special 
symplectic basis (although the equations may take a more 
`suggestive' form), as this only corresponds to taking linear 
combinations. Although we assumed the existence of the 
holomorphic prepotential, the above variational principle can 
also be formulated in a basis where such a prepotential does not 
exist, but for the purpose of this paper this feature is not 
important.  The components of $\Pi$ comprise $2N_V+2$ complex 
quantities, but only $N_V+1$ of them are independent (as the others are 
determined in terms of the prepotential). So generically, the 
above equation fixes $\Pi$ in terms of $p^I$ and $q_J$.
Before considering an explicit example, we note the following
convenient relations, which follow from (\ref{ycond}) for $p^0$
and $p^A$, 
\beqa
(z^A-\bar z^A) Y^0 = i (p^A -p^0\bar z^A)\,.\label{pacond}
\eeqa

As an example, consider the following cubic prepotential
\be
F(Y) = -b \,\frac{Y^1 Y^2 Y^3}{Y^0} + a \,\frac{ (Y^3)^3 }{Y^0} \;\;.
\label{prepotstu}
\eq
The solution to (\ref{ycond}) 
for a general magnetic/electric charge vector $(p^I,q_I)$
where $I=0, 1,2,3$, reads
\beqa
Y^0 &\! =\! & \frac{p^3 +ip^0 \bar{U} }{U + \bar{U}}\;,\qquad 
Y^1 = - \frac{Y^0}{ p^3 +i p^0 \bar{U} } \Big(-ip^1 \bar{U} +
\frac{q_2}{b} \Big)\;, 
\nonumber\\
Y^2 &\!=\!& - \frac{Y^0}{ p^3 +i p^0 \bar{U} } \Big(-ip^2 \bar{U}
+ \frac{q_1}{b} \Big) 
\;,\qquad 
Y^3 = iU\, Y^0\;,
\eeqa
where $U$ is 
determined by the following equation
\beqa
q_0 -i q_3 \bar{U} &=& {b\over  p^3 +i p^0 \bar{U}} \,\Big(-ip^1 \bar{U} 
+ \frac{q_2}{b}\Big)\Big(-ip^2 \bar{U} + \frac{q_1}{b}\Big)
 \nonumber \\
&&+ a\, p^3( U^2 + 2 U \bar{U} - 2 \bar{U}^2)
+i a \,p^0 U \bar{U}(  U + 2 \bar{U} ) \;\;\;.
\label{EqforU}
\eeqa
The entropy can be determined as a function of $U$, by
making use of (\ref{ADMKP}), 
\be
{{\cal S}\over \pi} = \bigg\vert {U+\bar U\over p^3-ip^0U}\bigg\vert^2\; 
\bigg\{{(b\, p^1p^3 + q_2p^0) (b\, p^2p^3 + q_1p^0)\over b}-a 
\bigg\}\;.
\eq
What remains is to solve (\ref{EqforU}). 
For the case $a=0$, $b=1$, this is a quadratic equation for $U$ with 
solution
\be
U= i{q_0p^0 +q_1p^1+q_2p^2-q_3p^3\over 2(q_3p^0 +p^1p^2)} \pm 
 \sqrt{ {q_1q_2 -q_0p^3\over q_3p^0 +p^1p^2 } -
{(q_0p^0 +q_1p^1+q_2p^2-q_3p^3)^2 \over 4(q_3p^0 +p^1p^2 )^2} }\;.
\eq
These solutions with the corresponding value for the entropy can 
be compared to previous results \cite{BKRSW,KSW,CLM} (for the 
results of the second and third work this comparison requires 
a conversion to the appropriate symplectic basis).


\subsection{Perturbative entropy formula for heterotic string 
compactifications} 

The classical entropy formula for $N=2$ supersymmetric heterotic string 
compactifications has been derived in the perturbative 
string basis \cite{KSW,CLM}. It was shown to be invariant (as 
should be expected) under 
the target-space duality group, which, at the classical level, 
is just equal to $SO(2,N_V-1)$. The entropy was also constructed in the 
symplectic basis corresponding to the first term in 
(\ref{hetprepot}) in \cite{BKRSW}.  

In this section we derive the entropy formula, but now to all orders of 
string perturbation theory. It reads
\be
{{\cal S}\over \pi} = {8\pi\over g_{\rm pert}^2}\,\bigg\vert_{\rm 
hor} \;
(p^0q_1 + p^a\eta_{ab}p^b)  \;\;.
\label{entdem}
\eq
The perturbative string coupling depends on the values of the 
dilaton field and the moduli at the horizon. 
The charges $p$ and $q$ refer to the
magnetic/electric charges as defined in the symplectic basis
associated with (\ref{hetprepot}). This is, however, not the basis
defined by perturbative string theory, where the magnetic charges
are equal to $N^I= (p^0, q_1, p^2,\ldots)$. These magnetic charges
transform linearly under target-space duality transformations,
\be
N^I\to \hat U^I{}_J\,N^J \;,
\eq
where the matrix $\hat U$ belongs to a subgroup of 
$SO(2,N_V-1,{\bf Z})$. 
In terms of the string basis we find that we are dealing with an
invariant under these transformations \cite{Ceresole,DKLL,CLM}
\be
p^0q_1 + p^a \eta_{ab} p^b= N^0 N^1 + N^a \eta_{ab} N^b \equiv
\langle N,N\rangle \;\;.
\eq
Thus, in the perturbative string basis, eq. (\ref{entdem}) reads
\beqa
{{\cal S}\over\pi}= {8\pi\over g_{\rm pert}^2}\,\bigg\vert_{\rm 
hor} \;  \langle N,N\rangle  \;.
\label{enthetstr}
\eeqa
Due to nonrenormalization theorems, this result is true to all 
orders in perturbation theory and takes precisely  the same form 
as the classical entropy  
formula \cite{CLM}, with the tree-level coupling
constant replaced by its full perturbative value. As the latter is 
invariant under target-space duality \cite{DKLL}, the 
perturbative entropy formula is invariant under 
target-space duality.  In fact, since  $\langle N,N\rangle$ 
is invariant under target-space duality transformations, whereas 
the dilaton is not, at least not beyond the classical level, 
it was natural to expect that the corrected entropy formula
should be given by the tree-level formula with the dilaton replaced by 
some one-loop target-space duality invariant object. It is 
gratifying to see that this object is precisely the true 
loop-counting parameter of heterotic string theory.

Let us now show that (\ref{entdem}) indeed holds.  
Inserting eqs. (\ref{Kloop}) into
eq. (\ref{ADMKP}) yields 
\beqa
{{\cal S}\over\pi}= (S+\bar S +V_{GS})\,|Y^0|^2\, 
(T^a + {\bar T}^a)\eta_{ab}(T^b + {\bar T}^b) \;.\label{moneloopa}
\eeqa
Using that $T^a = - i z^a$, and
inserting (\ref{pacond}) and its complex conjugate into (\ref{moneloopa})
yields
\be
|Y^0|^2\, 
(T^a + {\bar T}^a)\eta_{ab}(T^b + {\bar T}^b)       
=  (p^a +i p^0 {\bar T}^a) \eta_{ab}(p^b - ip^0 T^b) 
\;\;.\label{one}
\eq
On the other hand, it follows from (\ref{ycond}) and (\ref{hetprepot}) 
that 
\be
F_1(Y)  - {\bar F}_1(\bar Y) = -\frac{Y^a \eta_{ab}  Y^b}{Y^0}
 + \frac{{\bar  Y}^a \eta_{ab}  {\bar Y}^b}{{\bar Y}^0}
= i q_1 \;\;.
\eq
Using that ${\bar Y}^a = Y^a - i p^a$, we obtain 
\beqa 
F_1(Y) - {\bar F}_1(\bar Y) &=& -\frac{Y^a \eta_{ab}  Y^b}{Y^0}
 + \frac{{\bar  Y}^a \eta_{ab}  (Y^b- i p^b)}{{\bar Y}^0} \nonumber\\
&=& - iT^a \eta_{ab} Y^b - i{\bar T}^a \eta_{ab} (Y^b - i p^b) 
\nonumber\\ 
&=& 
 Y^0 (T^a + {\bar T}^a) \eta_{ab} T^b -  {\bar T}^a \eta_{ab} 
p^b=iq_1 \;\;.
\eeqa
Using once more (\ref{pacond}), we establish 
\be
 p^0 q_1 + p^a \eta_{ab} p^b =  (p^a +ip^0 {\bar T}^a) 
\eta_{ab}(p^b - ip^0 T^b) \;\;. \label{two}
\eq
Combining (\ref{moneloopa}), (\ref{one}) and (\ref{two}) and 
using the expression for the perturbative string-coupling 
constant (\ref{oneloopc}) yields the desired result (\ref{entdem}).


\subsection{The axion-free case}

Axion-free solutions are solutions with ${\rm Re} \; z^a =0$. For 
these solutions (\ref{pacond}) takes the form 
\be
z^A(2Y^0 -ip^0) = i p^A\,.
\eq
First let us assume that $2Y^0-i p^0= Y^0+\bar Y^0= 
\lambda\not=0$. In that case we  
easily derive the following result for the $Y^I$, 
\be
Y^0={\textstyle{1\over 2}} (\lambda + ip^0)\,,\qquad Y^A= ip^A 
\,{\lambda+ip^0\over 2\lambda}\,.
\eq
Consider the second set of equations (\ref{ycond}) 
applied to an arbitrary 
prepotential of the heterotic/type-II form,
\be
F(Y)= {d_{ABC}\,Y^AY^BY^C\over Y^0} + ic (Y^0)^2\,,
\eq
where $c$ is a real constant. In principle we could allow 
additional quadratic terms, which would still be explicitly 
solvable, at least for axion-free solutions. Arbitrary quadratic 
terms with {\em real}  coefficients can be easily incorporated by 
making use of  
suitable symplectic reparametrization. This will be discussed in 
the next subsection. For the case above 
(\ref{ycond}) now yields the following equations,
\be
q_0 = {d_{ABC}\,p^Ap^Bp^C\over \lambda^2}+ 2c\l \,,\qquad 
q_A = -{3p^0\over \lambda^2} \;d_{ABC}\,p^Bp^C\,, \label{nonax1}
\label{q0A}
\eq
leading to the condition
\be
3p^0q_0 + p^Aq_A=6c\l\,p^0\,.
\eq
Observe that the first condition (\ref{nonax1}) can only be 
satisfied for $(q_0-2\l)\,  d_{ABC} p^A p^B p^C > 0$. 
The entropy can be computed from (\ref{ADMKP}) and reads 
\be
{{\cal S}\over \pi} = -2 (q_0 - 2c\l) \bigg[\lambda + {(p^0)^2\over 
\lambda}\bigg] \;\;. 
\eq
For $cp^0\not=0$ we can express $\l$ in terms of the charges,
\be
\l = {3p^0q_0 + p^Aq_A\over 6c\,p^0} \,,
\eq
On the other hand, when $cp^0=0$ we have a constraint on the 
charges,
\be
3p^0q_0 + p^Aq_A= 0\,.
\eq
For $c=0$ and $q_0\not=0$ we can express $\lambda$ as
\be
\lambda= \pm \sqrt{ d_{ABC}\,p^Ap^Bp^C\over q_0} \,.
\label{LamQua}
\eq
Plugging this into (\ref{q0A}) one can express the charges
$q_A$ in terms of the remaining ones,  
$q_0, p^0, p^A$. Positivity of the entropy  
requires $q_0 \lambda < 0$. In the following we choose the moduli $z^A$ to
live on the upper-half plane Im~$z^A>0$ and for convenience we
restrict ourselves to charges with $q_0 < 0$ and
$p^A > 0$. Then the moduli $z^A$ take the form
\be
z^A= i\, p^A\, \sqrt{q_0\over d_{ABC}\,p^Ap^Bp^C}\,. \label{axfreemod}
\eq
As a special case, consider the non-axionic solution (\ref{q0A})
with $c=p^0=0$ and, consequently, $q_A=0$.
This constitutes a solution with only $N_V+1$
independent, non-vanishing charges, which we take to satisfy 
$p^A>0$, $q_0<0$, for definiteness.
The entropy is given by 
\begin{equation}
{{\cal S}\over\pi}=
2 \sqrt {q_0 d_{ABC} p^Ap^Bp^C}\label{cubicmadm} \;,
\end{equation}
and the moduli are given in (\ref{axfreemod}).
In particular, for the cubic prepotential (\ref{prepotstu}) we
find that 
\beqa
z^1 = {p^1}\,{z^3\over p^3}\,,\qquad z^2 = p^2\,{z^3\over  
p^3}\,,
\qquad 
z^3 = i \sqrt{ \frac{q_0 p^3}{-b\, p^1 p^2 + a (p^3)^2 } }\,,
 \label{onelooputs}
\eeqa
as well as 
\be 
{{\cal S}\over\pi}=2 \sqrt{ - q_0 (b \, p^1 p^2 p^3 - a (p^3)^3 ) }
\;.
\label{BPSmass}
\eq
For the values
$b=-3a=1$, the cubic prepotential (\ref{prepotstu}) describes
the one-loop corrected heterotic prepotential (\ref{STUprep}) 
of the $S$-$T$-$U$ model in the decompactification limit 
${\rm Re } \; S > {\rm Re}\; T > {\rm Re }\;  U  \rightarrow \infty$. 
Consistency of this limit
requires the following ordering of the absolute values of the charges:
$-q_0 \gg p^1 > p^2 > p^3 \gg 0$.
It will be shown in section 4 that this hierarchy of
charges also guarantees the suppression of $\alpha'$
corrections.

Also note that the solution (\ref{LamQua})
we found for the case
$c p^{0}=0$ is a good approximate solution for the general case
$c \not=0$ (with general $p^{0}$). 
Recalling that $c= \frac{\chi \zeta(3)}{16 \pi^{3}}$,
which is of order 1 for typical Calabi-Yau Euler numbers $\chi$
with $|\chi| \leq 1000$,
we expect that the constant term
in the prepotential will only give a small contribution when
the moduli are large. Comparing the exact solution for $\lambda$
in the case $c\not=0$ to the solution (\ref{LamQua}) one can
show that both differ by terms of order $\sqrt{ \frac{c^{2}
d_{ABC} p^{A} p^{B} p^{C}}{q_{0}^{3}}  } $, which is small
for $|q_{0}| \gg |p^{A}|$, i. e. for large moduli.

The second class of solutions corresponds to 
$Y^0={1\over 2}ip^0$, 
which implies (for finite $z^A$) 
that all the $p^A$ must vanish. 
Now the stabilization equations
(\ref{ycond}) imply that
\be
q_A= 3p^0 \,d_{ABC}\,z^Bz^C\,,\qquad q_0 =  0\,.
\eq
Hence the only nonzero charges are $p^0$ and (some of) the $q_A$. The above 
$N_V$ quadratic equations for the $N_V$ purely imaginary parameters $z^A$
can usually be solved straightforwardly. Note that there is no
dependence on the constant term $c$ in this case, 
because $Y^{0}$ is purely imaginary.

To demonstrate this second solution we reconsider the 
prepotential corresponding to (\ref{prepotstu}). The equations for 
$q_A$ take the form
\be
q_1= b\, p^0\,TU\,,\qquad q_2 = b\,p^0\,SU \,,\qquad q_3= b\,p^0\,  
ST  -3a\,p^0 \, U^2\,,
\eq
with $S$, $T$, $U$ real. 
These solutions can be solved for $S$, $T$ and $U$, 
\beqa
&&2\sqrt{b\,p^0q_1\over q_2} \,S=2\sqrt{b\,p^0q_2\over q_1} \,T = 
\sqrt{q_3 + 2\sqrt{-3a\,q_1q_2\over b}} + \sqrt{q_3 - 
2\sqrt{-3a\,q_1q_2\over b}} \;,\nonumber \\
&& 2\sqrt{-3a\,p^0}\,U= \sqrt{q_3 + 2\sqrt{-3a\,q_1q_2\over b}}- 
\sqrt{q_3 - 2\sqrt{-3a\,q_1q_2\over b}} \;.
\eeqa 
The charges and the coefficients $a$ and $b$ must be chosen such 
that $S$, $T$ and $U$ are positive.


\subsection{The entropy formula in type-II compactifications}

The entropy formula for extreme black holes in type-II 
compactifications will depend 
on electric and magnetic charges as well as on topological data
of the Calabi--Yau manifold, on which one has compactified the 
type-II string theory.  The  topological data 
appearing in the prepotential are the
classical intersection numbers $C_{ABC}$
as well as the expansion coefficients ${c_2 \cdot J_A}$ of the 
second Chern class $c_{2}$ of the three-fold, which were defined
in section 2.3. 
These data are related to the (real) coefficients $d_{ABC}$, $c$ and 
$W_{0A}$ of the associated prepotential,
\beqa
F(Y)={ d_{ABC}  \,Y^A Y^B  Y^C\over  Y^0}  + W_{0A} \, Y^0 Y^A 
+ i c (Y^{0})^{2}
\,
,
\label{linII}
\eeqa
by $ d_{ABC} = - {1 \over 6} C_{ABC}$ and 
$ W_{0A} = \frac{c_2\cdot J_A}{24}$.

For extreme black holes based on the prepotential (\ref{linII}), the
entropy formula will generically be given by
\beqa
{{\cal S}\over \pi} = \vert Z_{\rm hor}\vert^2 = 
{\textstyle{1\over 4}}\,A \left ((p^I ,
q_I ), C_{ABC}, c_2 \cdot J_A, \chi  \right) \;\;.
\eeqa

As is well known, quadratic polynomials with real coefficients 
can be introduced into any $N=2$ prepotential by a suitable 
symplectic reparametrization. So the above case (\ref{linII}) is 
covered by our previous analysis, provided we perform the 
corresponding symplectic rotation on the associated charges, 
\begin{equation}
 \pmatrix{\tilde{p}^I \cr \tilde{q}_I}  =
 \pmatrix{1 &0 \cr  W & 1}  \pmatrix{p^I \cr q_I} \;\;,
\end{equation}
where $W_{AB} = W_{00} =0 $. Note that a non--vanishing $W_{00}$
would not allow us to eliminate the term $i c (Y^{0})^{2}$ in the
prepotential, because $W_{00}$ must be real, wheras $ic$ is imaginary.
More general theta shifts with $W_{AB} \not=0,$ $W_{00} \not=0$
would generate quadratic and constant terms in $Y^{0}$ with
real coefficents. We will discard these terms, because they don't have
a topological interpretation.

In the following the electric and magnetic charges
of the former solution are denoted by $\tilde q_I$ and $\tilde p^I$, 
respectively, whereas the  electric and magnetic charges of the latter are
denoted by $q_I$ and $p^I$.
Note that this symplectic transformation induces a shift to the 
theta angles and thus a corresponding shift of the electric 
charges \cite{witteff}.  Thus, it follows that the 
entropy for the former solution can be computed from the entropy for 
the latter by performing the above substitution of the electric charges.

Consider, for instance, the axion-free solution (\ref{cubicmadm}) discussed
in the previous subsection, based on the cubic
prepotential (\ref{dfunction}), with $p^0=q_A=0$ and setting $c=0$.
Then we have for the symplectically transformed solution
that
\begin{equation}
 q_0 = \tilde{q}_0 - W_{0A} \tilde{p}^A \;\;,
\end{equation}
and for its entropy that 
\begin{equation}
{{\cal S}\over \pi} = 2  \sqrt { (\tilde{q}_0 - W_{0A} 
\tilde{p}^A) d_{BCD} \tilde p^B\tilde p^C\tilde p^D} \;\;.
\end{equation}
Thus, we can in particular set $\tilde{q}_0 = 0$, that is, we 
have a solution that is determined by magnetic charges $\tilde 
p^A$ only, which is non-singular and has non-vanishing entropy
 \begin{equation}
{{\cal S}\over \pi}= 
2 \sqrt { - ( W_{0A} \tilde{p}^A) d_{BCD} \tilde p^B\tilde p^C\tilde p^D}
\;\;.
\end{equation}
In the effective action, the term proportional to $W$ in 
(\ref{linII}) manifests itself in the presence of the additional 
term in the action 
\begin{equation}
\delta S \sim \int W_{A0} \, F^{A} \wedge F^{0} \ .
\end{equation}
Since $F^0$ is an electric gauge field and $F^A$ is a magnetic
monopole field, this integral is non-vanishing.


\section{Relation to higher-dimensional geometries}

\setcounter{equation}{0}

The black-hole solutions discussed so far appeared in the context
of either a compactification of the heterotic string 
on $K3\times T_2$ or of the type-II string 
on a Calabi--Yau three-fold. Type-II string theory, on the other 
hand, is dual \cite{witt} to $M$-theory compactified
on $CY\times S_1$  \cite{ferr}.  
In this section we discuss how the black-hole 
geometries associated with (\ref{cubicmadm}) arise from a 
compactification of the higher-dimensional
spacetime, that is,  by a compactification of $M$-theory.
We focus on those black-hole geometries that can either be obtained by 
a type-II string compactification on a Calabi--Yau three-fold with
$h_{1,1}=3$, or that are associated with
the $S$-$T$-$U$ models on the heterotic side.

On the $M$-theory side, we can regard these black-hole solutions
as arising from 
compactifications of certain 11-dimensional solutions describing 
three intersecting $M$-5-branes with a boost along the
common string.  Let us first consider the simplest such 11-dimensional
solution, which can be compactified on a 6-dimensional torus,
 \cite{ts}:
\begin{equation} \label{m-metric1}
 \begin{array}{l}
  ds_{11}^2 = {1\over (H^1 H^2 H^3)^{\frac{1}{3}}} \Big[du \,dv + 
  H_0 \,du^2 + H^1 H^2 H^3 \,d\vec{x}^2 +  \\
  \qquad\quad +  H^1 (dy_1^2 + dy_2^2) + H^2 (dy_3^2 + dy_4^2) + H^3
   (dy_5^2 + dy_6^2)   \Big] \;.
\label{eltorus}
 \end{array}
\end{equation}
Here, the $H^1,\ H^2$ and $H^3$ parametrize the three 5-branes and they are
harmonic functions with respect to $\vec{x}$. The internal space is
spanned by the coordinates $y$. Each 5-brane wraps around a 4-cycle;
e.g.\ the $H^1$-5-brane around $(y_3 , y_4 , y_5 , y_6)$, and any two
4-cycles intersect each other in a 2-cycle. 

Next, let us look at more complicated 11-dimensional solutions
which can be compactified on Calabi--Yau three-folds. 
For a generic Calabi--Yau three-fold, the intersection of three of the
4-cycles is determined by the classical intersection numbers
$C_{ABC}$.  This leads us to make the following ansatz
for the 11-dimensional metric, in analogy to (\ref{eltorus}),
\begin{equation}
  ds_{11}^2 = \frac{1}{({1 \over 6} C_{ABC} H^A H^B H^C)^{\frac{1}{3}}} 
  \Big[du \,dv + H_0 \,du^2 + {\textstyle{1 \over 6}} C_{ABC} H^A H^B 
H^C \,d\vec{x}^2 +   H^A \omega_A \Big] \;,
\label{eldabc}
\end{equation}
where $\omega_A$ ($A=1,2,3$) are the 2-dimensional line elements, which
correspond to the intersection of two of the 4-cycles. 
Below, we will fix the harmonic functions $H^A$ for the 
solution (\ref{cubicmadm}) in the double extreme limit.

After compactifying the internal coordinates in (\ref{eldabc}),
we obtain a magnetic
string solution in $D=5$ dimensions.
Similarly to the extreme Reissner-Nordstrom black hole in 
$D=4$ dimensions, this magnetic solution
has a non-singular horizon with the asymptotic geometry $AdS_3
\times S_2$ \cite{ho/to}.  In order to obtain a 
regular solution in $D=4$ dimensions as
well, we first have to perform a boost along this string (parameterized by
$H_0$), which will keep the compactification radius $G_{uu}$ finite
everywhere. This boost will induce momentum modes propagating along the
magnetic string. Turning off these modes has the consequence that this
radius shrinks to zero size on the horizon and that the solution becomes
singular.  Thus, performing the boost adds one electric charge 
to the three magnetic charges. Then, all the 
radii of the Calabi--Yau 2- and 4-cycles as well as of the string will
also stay finite
on the horizon.  The resulting 4-dimensional metric defines an extreme
Reissner-Nordstrom geometry given by
\begin{equation} \label{4dmetricm}
 ds_4^2 = - \frac{1}{\sqrt{ - {1 \over 6 } H^0 \, C_{ABC} H^A H^B
     H^C}} \,dt^2 + 
  \sqrt{- {\textstyle{1 \over 6}} H_0 \, C_{ABC} H^A H^B H^C } \, 
d\vec{x}^2 \;.
\end{equation}

Next, let us consider the dual heterotic string solution 
with fields $S$, $T$ and $U$.
This will allow us to determine the harmonic functions $H^A$.
We will 
restrict ourselves to the classical solution, that is to (\ref{BPSmass})
with $b=1, a=0$. 

First, we will have to change the symplectic basis.  That is, we 
will have to go from the basis corresponding to (\ref{hetprepot}) 
to the perturbative basis preferred by the heterotic string. 
This requires a symplectic reparametrization, after which $p^1$ is no 
longer a magnetic, but an electric charge: $p^1\rightarrow 
-q_1$. Hence in the heterotic string basis the  
solution is now characterised by 2
magnetic ($p^2,p^3$) and 2 electric ($q_0,q_1$) charges. The
classical 
$S$-$T$-$U$ black hole can then be obtained from the
6-dimensional solution  \cite{cv/ts}
\begin{equation}
 ds_6^2 = {1 \over H_1} \Big( du \,dv + H_0 \,du^2 \Big) +
 H_2 \Big({1 \over H_3} (dx_4 + \vec{V} d\vec{x})^2 + H_3 \,
d\vec{x}^2  \Big) 
\end{equation}
($\epsilon_{ijk} \partial_j V_k = \partial_i H_3$).
It describes a fundamental string lying in a solitonic 5-brane. Again, in 
order to
keep the compactification radii finite, we need to perform a boost along
the string and put a Taub-NUT soliton in the 
transversal space. From the resulting solution,
we can immediately read off the $S,T$ and $U$
fields. By compactifying over $u$ and $x_4$, we obtain for the internal 
metric that 
\begin{equation}
 G_{rs} = \pmatrix{ {H_0 \over H_1} & 0 \cr 0 & {H_2 \over H_3}}
  = {H_2 \over H_3} \pmatrix{ ({\rm Re} \,U)^2 & 0 \cr 0 & 1} \;,
\end{equation}
and thus we find for the scalar fields that
\begin{equation}
 \begin{array}{l}
 S =  e^{-2 \phi} =  e^{-2 \hat{\phi}} \sqrt{|G_{rs}|} = 
   \sqrt{{H_0 H_1 \over H_2 H_3}} \;, \\
 T =  \sqrt{|G_{rs}|} =   \sqrt{{H_0 H_2 \over H_1 H_3}} \qquad , \qquad 
 U =  \sqrt{{H_0 H_3 \over H_1 H_2}}  
\label{stuh}
 \end{array}
\end{equation}
($\hat{\phi}$ is the 6-dimensional dilaton).
In the double extreme limit, we have to fix the values of the
scalars at infinity so that they are constant everywhere. For the harmonic
functions this means that
\beqa \label{harmon}
&&H_0 = \sqrt{2} q_0 \Big(c + {1 \over r}\Big) \ , \qquad
H_1 = \sqrt{2} q_1 \Big(c + {1 \over r}\Big) \ ,\\ 
&&H_2 = \sqrt{2} p^2 \Big(c + {1 \over r}\Big) \ ,\qquad 
H_3 = \sqrt{2} p^3 \Big(c + {1 \over r}\Big) \;, \nonumber
\eeqa
with $c^{-4} = 4 q_0 q_1 p^2 p^4$ (in order
to obtain asymptotic Minkowski geometry in
$D=4$). The limit of large $q_0$ now has the consequence that the
boost or momentum along the string becomes large.  Hence, this
direction decompactifies and we obtain the 5-dimensional string solution.
The metric (in the Einstein frame) is in this case again given by
(\ref{4dmetricm}) with $C_{123} = 6$
as the only non-vanishing element.

We can now insert the harmonic functions (\ref{harmon}) 
into the metric (\ref{4dmetricm}) with general coefficients $C_{ABC}$.
In this way we precisely recover 
the metric (\ref{4dmetric}).  Note that in
our notation $-q_0 \, {1 \over 6} C_{ABC} p^A p^B p^C >0$.  When
approaching the
horizon $r \rightarrow 0$, we obtain the Bertotti-Robertson geometry which
is non-singular ($AdS_2 \times S_2$) and restores all supersymmetries.
The radius of the $S_2$ is given by the mass and so the area of the
horizon is $A = 4\pi M^2 = 8 \pi \sqrt{-q_0 \, {1 \over 6} C_{ABC} p^C p^B
p^C}$. This is the metric in the Einstein frame. The string, however,
couples to the string-frame metric, which can easily be given on 
the heterotic side. Replacing $-{\textstyle{1\over 6}} C_{ABC}$ 
by $d_{ABC}$ and using the dilaton value $S= -i z^1$ with $z^1$ 
given by (\ref{axfreemod}), we find that 
\beqa
 ds^2_{str} &=& {\sqrt{q_0\, d_{ABC}\, p^A p^B p^C}
  \over |q_0 p^1|} \, ds^2 \\
& =&    - {1 \over |q_0 p^1|} \Big(c + {1 \over r}\Big)^{-2}\,  dt^2 -
  { d_{ABC} \,p^A p^B p^C \over p^1 }  \Big(c + {1 \over 
r}\Big)^2 \,  d\vec{x}^2  \nonumber
\eeqa
(for $q_0 <0$ and $p^A >0$). This again has a throat geometry
for $r \rightarrow 0$
\begin{equation}
 ds^2_{str} \rightarrow - e^{2 \eta / R} \,dt^2 + d\eta^2
 + R^2 \,d\Omega^2  \qquad , \qquad R^2 = 
- { d_{ABC}\, p^A p^B p^C \over p^1}
\end{equation}
($r \sim \exp({\eta/ R})$).  Since the curvature has its maximum
inside the throat, we can keep higher curvature corrections ($\sim
{\cal O}(\alpha'))$ under control if the radius of the throat is
sufficiently large: $-d_{ABC}\, p^A p^B p^C \gg p^1$. This means
that sufficiently large magnetic charges ensure that all higher curvature
terms can be suppressed. 

\medskip


\section{Summary}

\setcounter{equation}{0}

Supersymmetric black holes provide us with a tool to probe the 
properties of the future fundamental theory which will describe  
non-perturbative quantum gravity.
This theory is expected to explain in a quantum-mechanical 
context the existence of all non-perturbative states, or 
solitons,  in string theory and in supergravity and also to
control the interaction between these states. Meanwhile, in the 
absence of such a theory, it is important to 
study supersymmetric black holes and their properties as the most
particular
representatives of the non-perturbative states of quantum 
gravity.  One of the remarkable property of all ($N=2,D=4$)
supersymmetric black holes is the topological nature
of the area of the black hole horizon in the sense that the area 
does not depend on the values of the moduli fields at spatial 
infinity \cite{ExtrN2BHs}. 
The explanation of the entropy via the counting of string states 
\cite{Microscopic} was thus established for a class of
black holes for which the entropy depends only on electric and 
magnetic charges. 
In this paper we have found various new area formulae for a class 
of $N=2, D=4$ supersymmetric theories.
The choice of the prepotentials is
motivated by various versions of string theory at the classical 
level as well as by string-loop corrections.

On the heterotic side  we have studied the prepotentials which 
include the contributions from string-loop corrections. 
At the perturbative level, we could prove that
the entropy takes precisely the same form as the tree-level
entropy \cite{KSW,CLM}, where the tree-level coupling constant $S+\bar S$
is replaced by the perturbative coupling constant, which 
originates entirely from one-loop effects and contains
the Green-Schwarz modification:
\begin{equation}
{{\cal S}\over\pi} = ( S + \bar S +  V_{GS })\Big\vert_{\rm 
hor}\;(p^0q_1 + p^a\eta_{ab}p^b).
\end{equation} 
Therefore, we confirmed the conjecture \cite{CLM} that
the string loops will affect the
area formula only via a perturbative modification of the string 
coupling.  
In case that the one-loop heterotic prepotential can be approximated
by a cubic polynomial, as it is true for large moduli values,
the one-loop string coupling can be explicitly expressed in terms
of the magnetic/electric charges for non-axionic black-hole 
solutions. For solutions that are not axion-free, the explicit 
expressions depend on solving some higher-order polynomial 
equation, as exhibited in section~3.1.

On the type-IIA side, our new area formulae imprint also the 
topological data of the Calabi-Yau 
manifold, in particular the intersection numbers $C_{ABC}$. The 
fact that this symmetric tensor enters the area formula for the 
five-dimensional black holes was known before \cite{FerKal1}.
However, in  five dimensions 
the area formula  is implicit, as one still has to minimize it in
the moduli space. Here, for the first time, we have found the 
area formulae of four-dimensional black holes which  
depend  on charges and on arbitrary intersection
numbers $C_{ABC}$. In addition, we have found an interesting 
dependence of the area on the second 
Chern class of the three-fold $c_2$ . Here we deal with the   
Witten-type shift \cite{witteff} 
of the electrical charge via magnetic charge in the presence of axions.
Finally the entropy will in general depend on the Euler number $\chi$ of the
Calabi-Yau three-fold, but this contribution is small compared to
the other effects we studied.

In the simplest case, when all the moduli $z^A$
are imaginary (the axion-free solution with $p^{0} = q_{A} = 0$
and $c=0$),  the entropy is given by
\begin{equation}
{\cal S}= 2 \pi
 \sqrt { \left(- {q}_0 + {c_2 \cdot J_{A} \over 24} {p}^A\right) 
{C_{BCD} \over 6}
p^B p^C p^D} \;.
\label{new}\end{equation}
This formula reproduces some previously known solutions, in 
particular for the $S$-$T$-$U$ black holes \cite{BKRSW},  where 
$C_{123}=6$ and  $c_2=0$, and where the entropy of
the simplest non-axion solution was found to be ${\cal S}= 2 \pi
\sqrt { |{q}_0 p^1 p^2 p^3|}$. 
One can now address the following
issue: which fundamental theory is capable of giving
a microscoping interpretation to (\ref{new})?

An interesting feature of the new area formulae is the 
possibility to relax some of the electric charges due to the 
above mentioned shift effect via $c_2 \cdot J_A$ terms. A simple 
example of such a relaxation is as follows. 
When applying the theta-angle shift to the 
$S$-$T$-$U$ black hole 
\cite{BKRSW} with one magnetic and three electric charges, which has the 
area formula $ {\cal S}= 2 \pi
 \sqrt { |{p}^0 q_1 q_2 q_3|}$, we obtain an
entropy formula which is non-vanishing for $q_1=q_2=q_3=0$, even though 
there now is only one magnetic charge present:
$ {\cal S}/\pi= 2 (p^0)^2
 \sqrt {  {1\over 24^3} | (c_2\cdot J_1) (c_2\cdot J_2)(c_2\cdot J_3)|}$.

In conclusion, we have found various qualitatively new features of
supersymmetric black holes in $N=2,D=4$ supergravity 
theories motivated by  string theory. 

\section*{Acknowledgements}
This work is supported by the European Commission TMR programme 
ERBFMRX-CT96-0045. The work of K.B. and T.M. is supported by DFG.  
The work of R.K. was  supported by  NSF grant PHY-9219345.  
She is grateful to the Institute of Physics of the
Humboldt University  Berlin for the hospitality.


\end{document}